\documentclass[aps,prl,twocolumn,superscriptaddress,showpacs]{revtex4}

\usepackage{subfigure}
\usepackage{graphicx}% Include figure files
\usepackage{bm}% bold math
\usepackage{amsmath}
\usepackage{epstopdf}

\newcommand{\STO}{SrTiO$_3$}
\newcommand{\LAO}{LaAlO$_3$}

\newcommand{\etal}{\textit{et al.}}

\begin{document}

\title{Anomalous magnetic ground state in \LAO/\STO~ interface probed by transport through nanowires}

\author{A.Ron}
\affiliation{Raymond and Beverly Sackler School of Physics and Astronomy, Tel-Aviv University, Tel Aviv, 69978, Israel}
\author{E.Maniv}
\affiliation{Raymond and Beverly Sackler School of Physics and Astronomy, Tel-Aviv University, Tel Aviv, 69978, Israel}
\author{D.Graf}
\affiliation{National High Magnetic Field Laboratory, Florida State University, Tallahassee, Florida 32310, USA}
\author{J.-H. Park}
\affiliation{National High Magnetic Field Laboratory, Florida State University, Tallahassee, Florida 32310, USA}
\author{Y.Dagan}
\affiliation{Raymond and Beverly Sackler School of Physics and Astronomy, Tel-Aviv University, Tel Aviv, 69978, Israel}

\begin{abstract}
Resistance as a function of temperature down to 20mK and magnetic fields up to 18T  for various carrier concentrations is measured for nanowires made from the \STO/\LAO~ interface using a hard mask shadow deposition technique. The narrow width of the wires (of the order of 50nm) allows us to separate out the magnetic effects from the dominant superconducting ones at low magnetic fields. At this regime hysteresis loops are observed along with the superconducting transition. From our data analysis we find that the magnetic order probed by the giant magnetoresistance (GMR) effect vanishes at $T_{Curie}=954 \pm 20 $ mK. This order is not a simple ferromagnetic state but consists of domains with opposite magnetization having a preferred in-plane orientation.
\end{abstract}


\maketitle
The nature of the magnetic state at the interface between \LAO~ and \STO~ and its coexistence with superconductivity has been at the focus of recent intense scientific research. Brinkman \etal~ found strong, sweep-rate dependent hysteretic behavior of the magneto-resistance \cite{brinkman2007magnetic}. Additional transport experiments suggested magnetic order at high magnetic fields \cite{shalom2009anisotropic, seri2009antisymmetric, flekser2012magnetotransport}. Transport in AFM written nanowires, has been attributed to a spin based mechanism related to emergent magnetism\cite{LeviAFMwiremagnetic}. Bert \etal \cite{bert2011direct} Dikin \etal~ \cite{dikin2011coexistence} and Li \etal~ \cite{li2011coexistence} suggested that superconductivity and ferromagnetism do coexist in the same temperature-magnetic field domain in the phase diagram but no evidence for spatial coexistence has been demonstrated. On the other hand, while torque magnetometry \cite{li2011coexistence} suggested a strong magnetic moment per site, of the order of $0.3\mu_B$, with $\mu_B$ the Bohr magneton, the scanning SQUID experiment \cite{bert2011direct} and $\beta$ NMR measurements\cite{PhysRevLett.109.257207} suggested weaker magnetization. X-ray magnetic circular dichroism and X-ray absorption spectroscopy suggested that unique magnetism reside on the titanium $d_{xy}$ orbital \cite{lee2013titanium}. Recently, we have shown that the conductance through a ballistic quantum wire has step height of $\frac{e^2}{h}$ \cite{RonQuantumwire} indicative of removal of spin degeneracy.
\par
On the theory side a mechanism for the coexistence of superconductivity on the homogeneous background of localized magnetic moments has been suggested \cite{michaeli2012superconducting}. Banerjee \etal~ suggested a spiral order to reconcile the discrepancy between the various magnetization probes \cite{banerjee2013ferromagnetic}. Ruhman \etal~ interpreted the magneto-transport properties as a competition between Kondo-screening at high carrier concentration and magnetism for low carrier concentration \cite{ruhman2013competition}.

\begin{figure}
 \includegraphics[width=1\hsize]{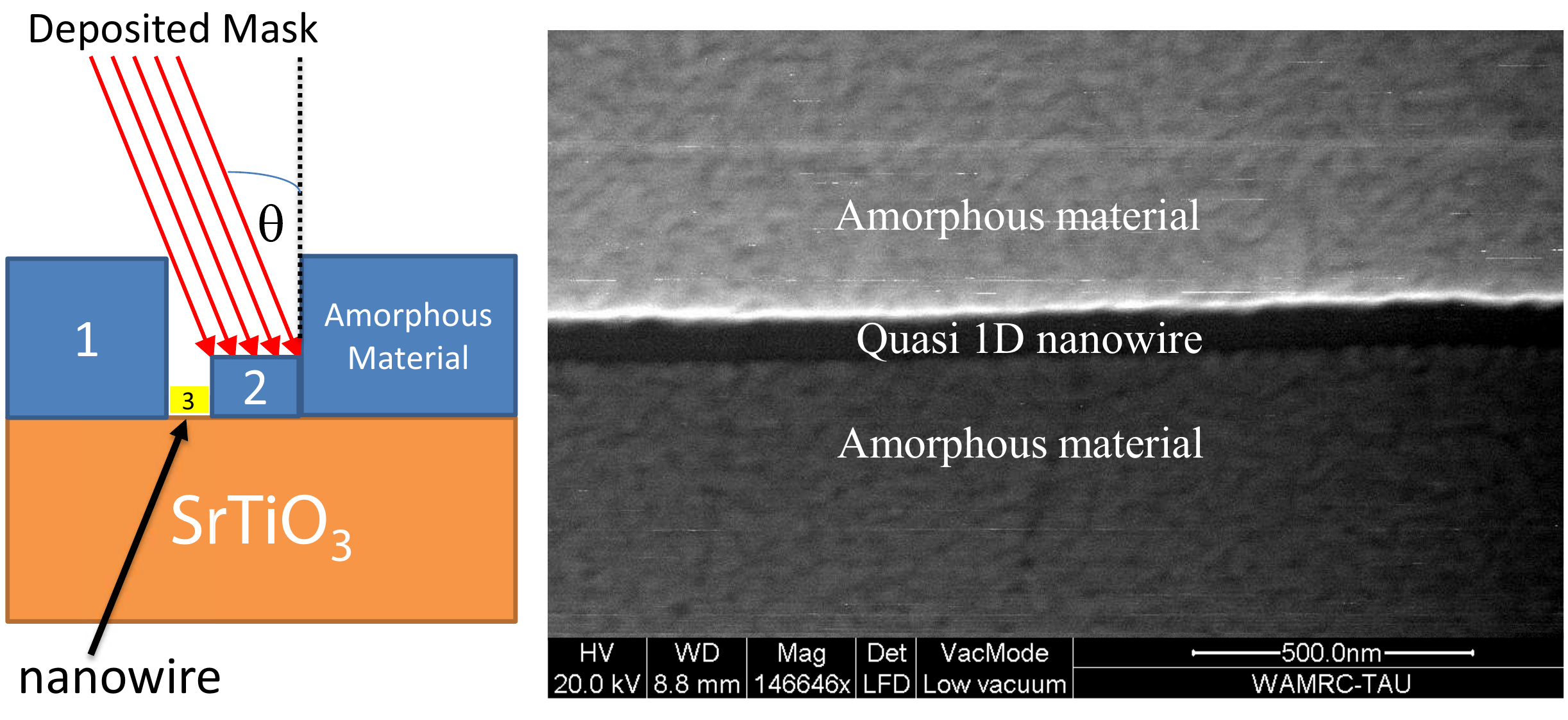}
  \caption {(color online).(Left panel) Fabrication method of the nanowires. On top of TiO$_2$ terminated \STO~ substrate an amorphous oxide layer number 1 with a thickness $d$ was deposited and patterned with standard optical lithography process. The substrate is then tilted to an angle $\theta$ and a second amorphous oxide layer is deposited. This results in a nano-trench with a width $w=d\tan(\theta)$. Finally an epitaxial \LAO~ layer is deposited as described in the text resulting in a conducting \LAO/\STO~ nanowire. Right panel: Scanning electron microscope image of such nanowire designed with $w=70nm$ consistent with the dimensions measured by the microscope.
}\label{Fig:1}
\end{figure}

\par
Fabricating nanostructures out of the \LAO/\STO~ interface is challenging \cite{rakhmilevitch2010phase, trisconenarrowbridge, chang2013quantum} since \STO~ is prone to spurious conductivity generated by oxygen vacancies. We have recently shown that the boundary between two nonconducting interfaces is a ballistic quantum wire \cite{RonQuantumwire}. Here we present a new method for fabricating nanowires. Measurements of such 50nm wide wires exhibit Shubnikov-de Haas oscillations (SdH) at high magnetic fields of the order of 10 T. The small width of the wires allowed us to separate out the magnetic contributions to the resistance from the dominant superconductivity and to observe hysteresis effects below the superconducting critical field H$_{c}$. The shape of the hysteresis and its angular dependence suggest an anomalous magnetic ground state consisting of adjacent magnetic domains aligned antiparallel.
\par
Epitaxial films of \LAO~ 10 unit cells thick are deposited using reflection high energy electron diffraction (RHEED) monitored pulsed laser deposition on atomically flat TiO$_2$ terminated \STO~(100) 0.5mm thick substrates in standard conditions, oxygen partial pressure of $1\cdot10^{-4}$ Torr and temperature of $780^oC$, as described in \cite{shalom2009anisotropic} followed by an annealing step at $400^oC$ and oxygen pressure of 200mTorr for one hour to minimize the contribution of oxygen vacancies to conductivity \cite{OxygenVacanciesConductivity} and magnetism\cite{OxygenVacanciesMagnetism} . Prior to the deposition of \LAO~ the samples were patterned with a hard mask of an amorphous oxide defining 5$\mu$m$\times50$nm nano-wires as depicted in Figure 1. The RHEED system was used to calibrate the deposition rate by performing a deposition at the above conditions on a large sample prior to that of the nanowires, after which a third RHEED monitored film is deposited to ensure that the calibration remained valid. A gold layer is evaporated as a back-gate. Ti-Au contacts were evaporated after Ar ion milling of the contact area. We use a wire bonder to connect voltage and current leads to each Ti/Au contact to eliminate the resistance of the leads, without eliminating the resistance between the metal and the nano-wires. All samples were cooled down in a $^3$He refrigerator, and few of them were also cooled down in a dilution refrigerator with a base temperature of 20mK. Magnetic fields as high as 18T were applied at various orientations. In the parallel orientation the field was either perpendicular to the current (transverse) or parallel to it (parallel). The presented sample resistances were measured using a Lakeshore 370 resistivity bridge with 3716L low resistance preamplifier and scanner. In order to distinguish magnetic hysteresis effects intrinsic to the sample from the remanent field of the superconducting magnet we also performed the measurements in a quenched magnet. In order to verify that the magnetic and superconducting phenomena observed are intrinsic to the samples the zero and low field measurements were also reproduced in three different dilution refrigerators equipped with different lock-in amplifiers and applied currents between 0.1 to 5nA. At least one of the measurement set-ups has been proven to show superconducting transition down to zero resistance in nanowires \cite{sternfeldnano}. In this letter we report data obtained from one sample these data were reproduced in 5 other samples fabricated on two different substrates.
\par

In Figure 2(a) we show the resistance as function of temperature. Clearly a superconducting transition is observed, however the resistance of the sample does not reach zero, this point will be addressed in the discussion. In Figure 2(b) we show the resistance of the sample as function of magnetic field applied perpendicular to the sample at 20 mK and $V_g=0 V$ (as grown state). SdH oscillations are observed at perpendicular fields greater than 10T  as shown in Figure 2(c). We use this quantum effect to characterize our samples. According to Luttinger's theorem, the period of the SdH oscillations is directly related to the two dimensional carrier density $n_{2D}$ by: $n_{2D}=eN_dF/h$
, where $N_d$ is the degeneracy. Ignoring any degeneracy, the obtained frequencies (see fast Fourier transform in Figure 2(d)) correspond to carrier densities of the order of $10^{12} cm^{-2}$ and can be modulated by application of gate voltage. These values are similar to those obtained for large samples \cite {shalom2010shubnikov}.
\par
In Figure 2(e) the magnetoresistance at 20mK for various gate voltages (going down from 31.2 V to 22.8 V) is shown. The resistance, magnetoresistance and superconducting properties are changing monotonically as reported for the two dimensional (2D) system in ref.\cite{shalom2010tuning}.
\par
%Figure 2
\begin{figure}
 \includegraphics[width=1\hsize]{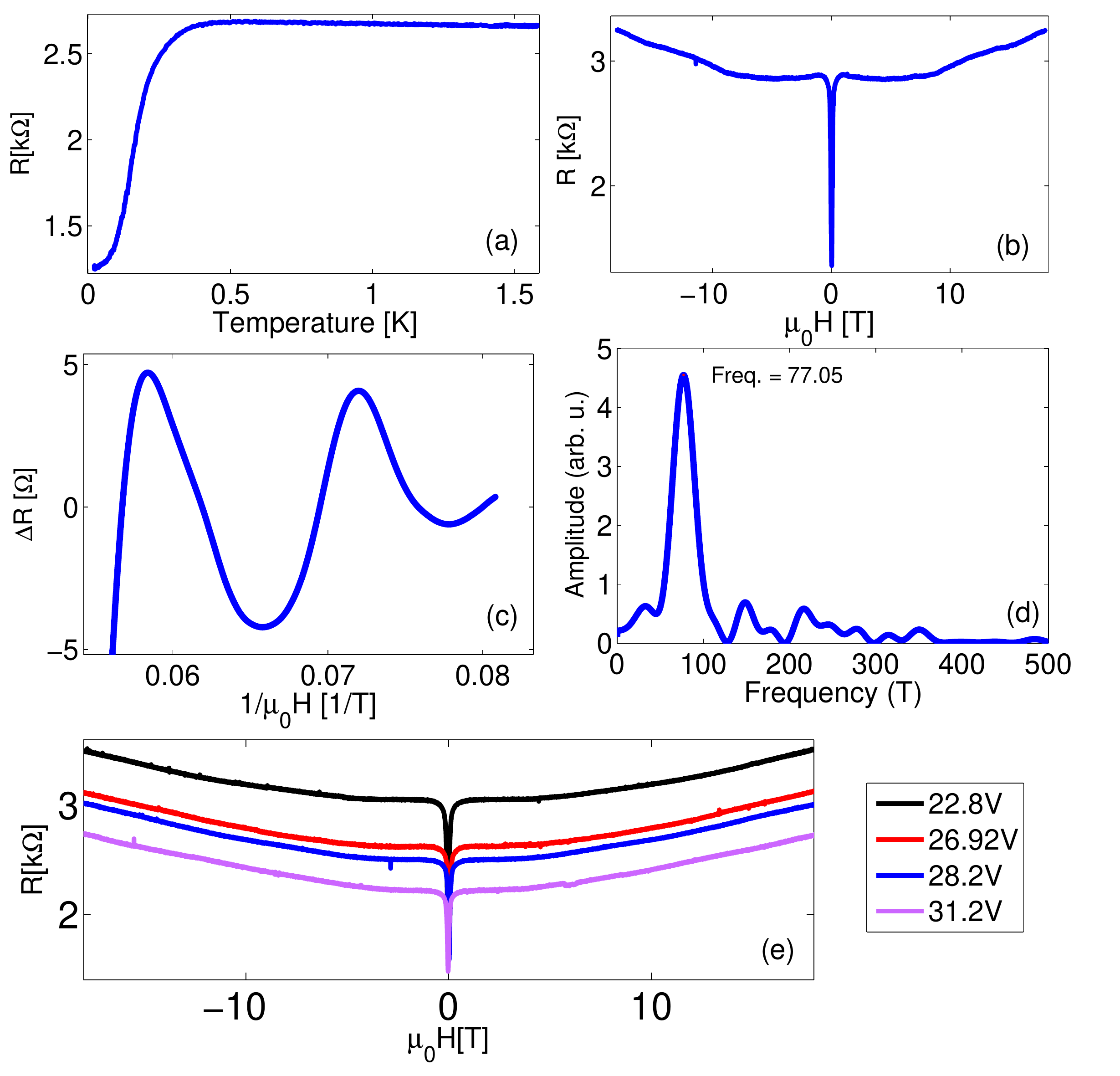}
  \caption {(color online).(a) Resistance as function of temperature at zero magnetic field and $V_g=0 V $ (as grown). (b) Resistance as a function of magnetic field at 20 mK and $V_g=0 V$ (as grown) (c) Focus on the Shubnikov-de Haas (SdH) oscillations plotted as function of inverse field after application of a low pass filter to reduce measurement noise. Periodic oscillations are observed.(d) A fast fourier transform of the SdH oscillations yielding a sharp peak at 77.05 T. (e) Resistance versus magnetic field for various gate voltages. The gate voltage is decreased from 31.2 V (bottom magenta curve ) to 22.8 V (top black curve)
}\label{Fig:1}
\end{figure}
\par

Figure 3 focuses on the low magnetic field dependence (up to $\sim$ 200 G) at 20mK for magnetic field applied perpendicular to the interface (Figure 3(a)), transverse (Figure 3(b)) and parallel to the wire (Figure 3(c)). As the field is swept back and fourth (blue curves and red curves) a negative hysteretic magnetoresistance is observed. For perpendicular field orientation, subtraction of the superconducting background (see caption) allows us a clearer observation of the magnetic effect. The curves are reproducible and their features are independent of field sweep rate (between the rates of 0.005 and 0.2 T/min).
\par
The magnetic features are similar to the giant magnetoresistance (GMR) effect reported in granular ferromagnetic materials \cite{chien1993giant}. The presence of a similar magnetic feature in all field orientations eliminates the anisotropic magnetoresistance (AMR) as an explanation for the effect. In this GMR description the low resistance state appears when the domains are aligned while they are not aligned for the high resistance state. This interpretation of our results is schematically depicted in Figure 3(d). The saturation field, $H_s$ is determined as the onset of the low resistance state.
\par
We note the hysteretic behavior of resistance in all three field directions. Such hysteretic behavior is expected for magnetic field applied along an easy axis \cite{stoner1948mechanism}. Vortices pinned to superconducting regions cannot be at the origin of the hysteresis since it is observed for the parallel field orientation where vortices do not exist. We can therefore conclude that the magnetization tends to align in the plane or along the crystal principle axis. A conspicuous difference between the in-plane and perpendicular field orientations is the abrupt transition from the high resistance to the low resistance states for the parallel direction. This suggests that the magnetic moment is pinned in the plane with a characteristic energy corresponding to a field of $\simeq 100$ Gauss. Furthermore a careful examination of our curves shows that the resistance returns to its zero field value before the magnetic field changes its direction (before reaching zero field). This is indicative of a non trivial magnetic ground state with magnetic domains aligned antiparallel.

%Figure 3

\begin{figure}
\centering
 \includegraphics[width=1\hsize]{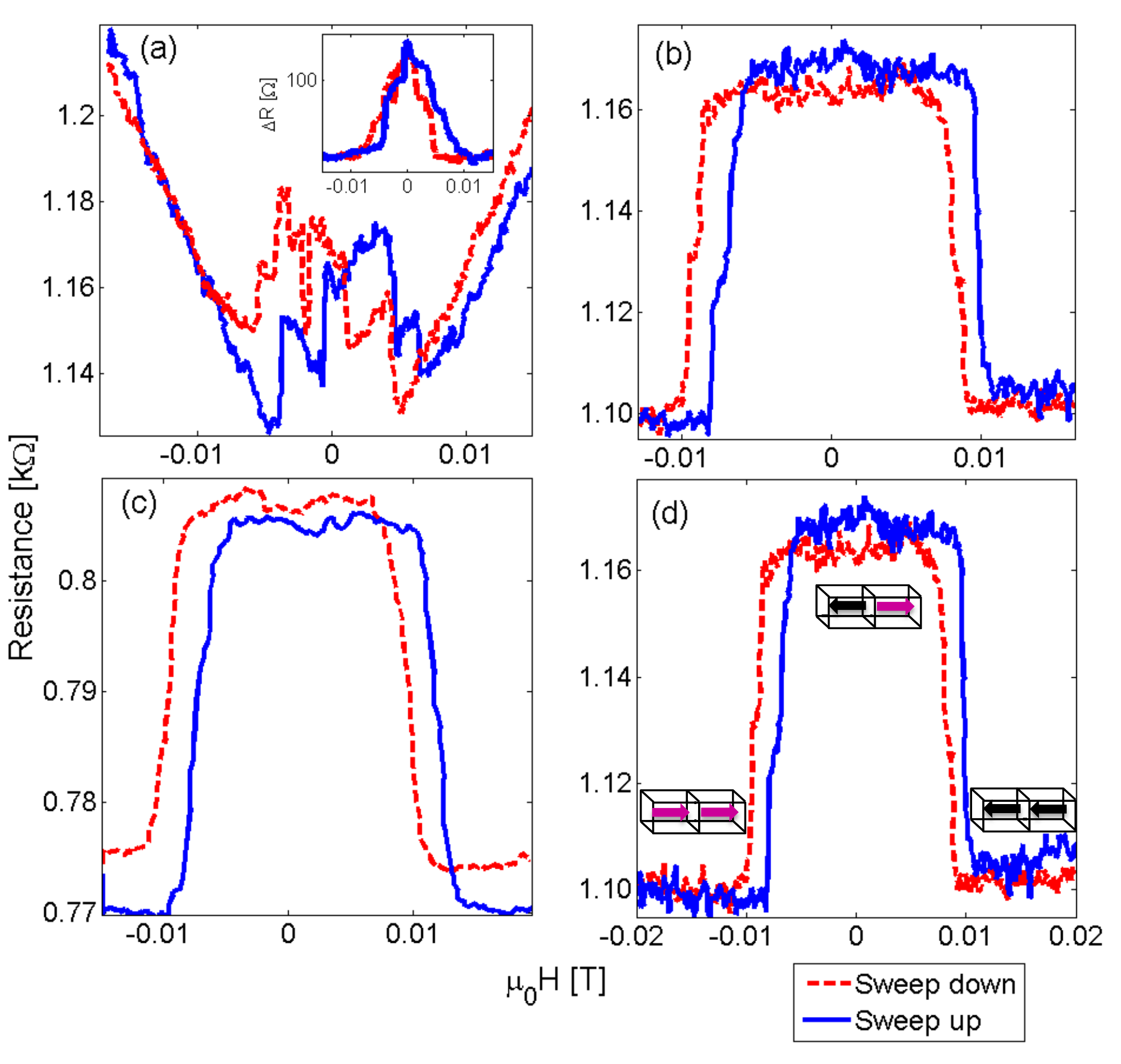}
  \caption{(color online). Resistance as function of magnetic fields for the low fields region at 20mK. Reproducible magnetic hysteresis curves are observed in (a) perpendicular (b) transverse  and (c) parallel  field orientations. Solid blue lines correspond to the curve plotted while increasing the field and the dashed red ones for decreasing it. Inset: The magnetic effect shown in (a) after subtraction of the superconducting background, a function of the form $a|H|+b$. (d) Schematic illustration of adjacent domain orientation in the various resistance states. The data are taken from (b).}\label{Fig:3}
\end{figure}
Figure 4 shows the resistance versus magnetic field for various orientations. We use these data to determine the angular dependence of $H_s$, which is plotted in Figure 4(b). As expected there is strong anisotropy between parallel and perpendicular field orientations. The anisotropy between parallel and transverse field directions strongly suggests that the magnetic effects come from the conducting region (the nanowire). It is important to note that application of gate voltage between -10 to 50 Volt did not change the saturation field neither in the perpendicular configuration nor in the transverse one. This result is in line with gate independent magnetization found by Kalisky \etal~ \cite{kalisky2012critical}.

%Figure 4
\begin{figure}
 \includegraphics[width=1\hsize]{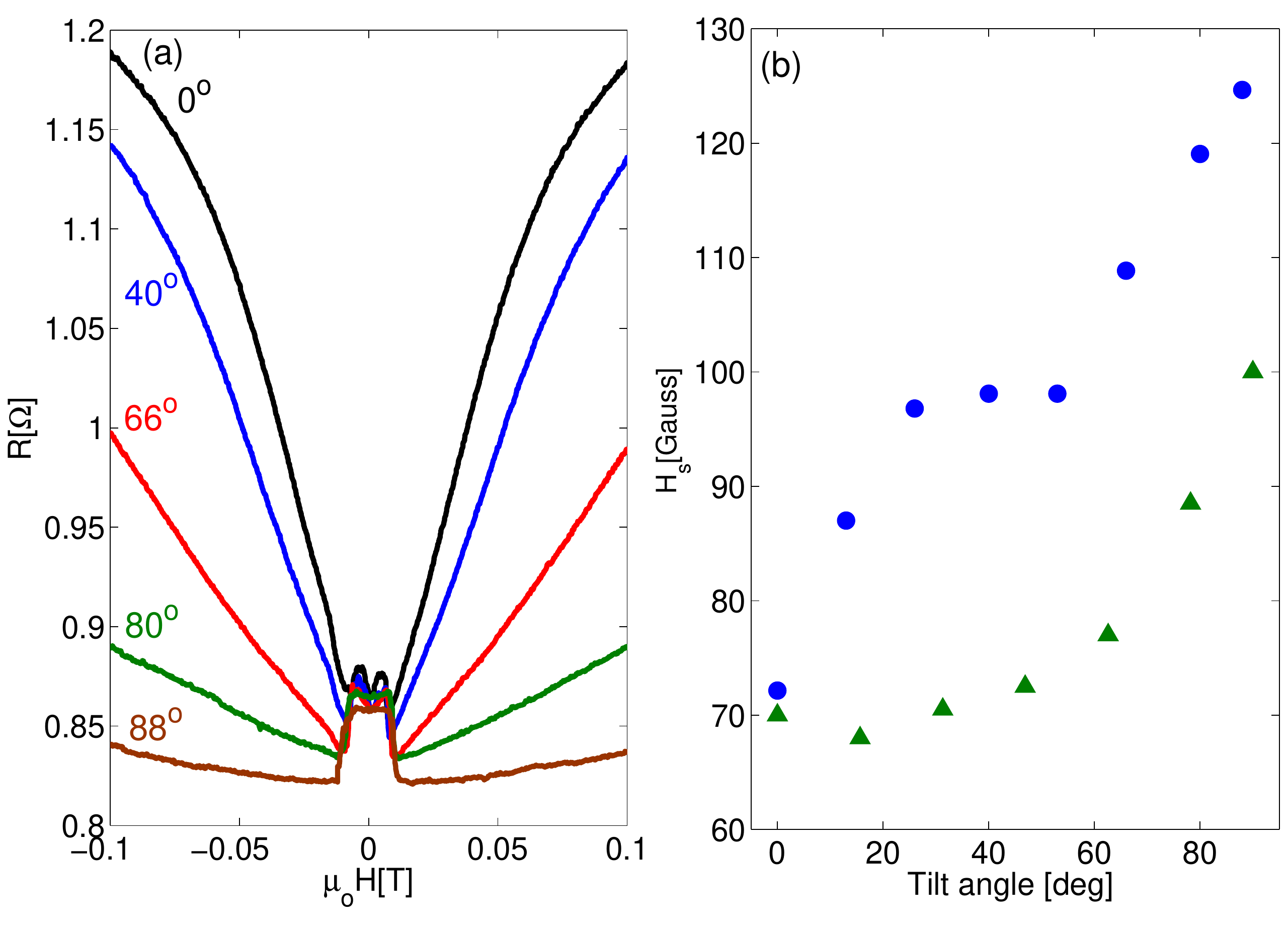}
  \caption{(color online). (a) Resistance versus magnetic field for various angles between the perpendicular to the interface and the applied magnetic field 90 degrees corresponds to the parallel field direction. (b) Saturation field for various angles. For the circles (triangles) 90 degrees corresponds to the parallel (transverse) field orientation.}\label{Fig:4}
\end{figure}
%Figure 5
\begin{figure}
 \includegraphics[width=1\hsize]{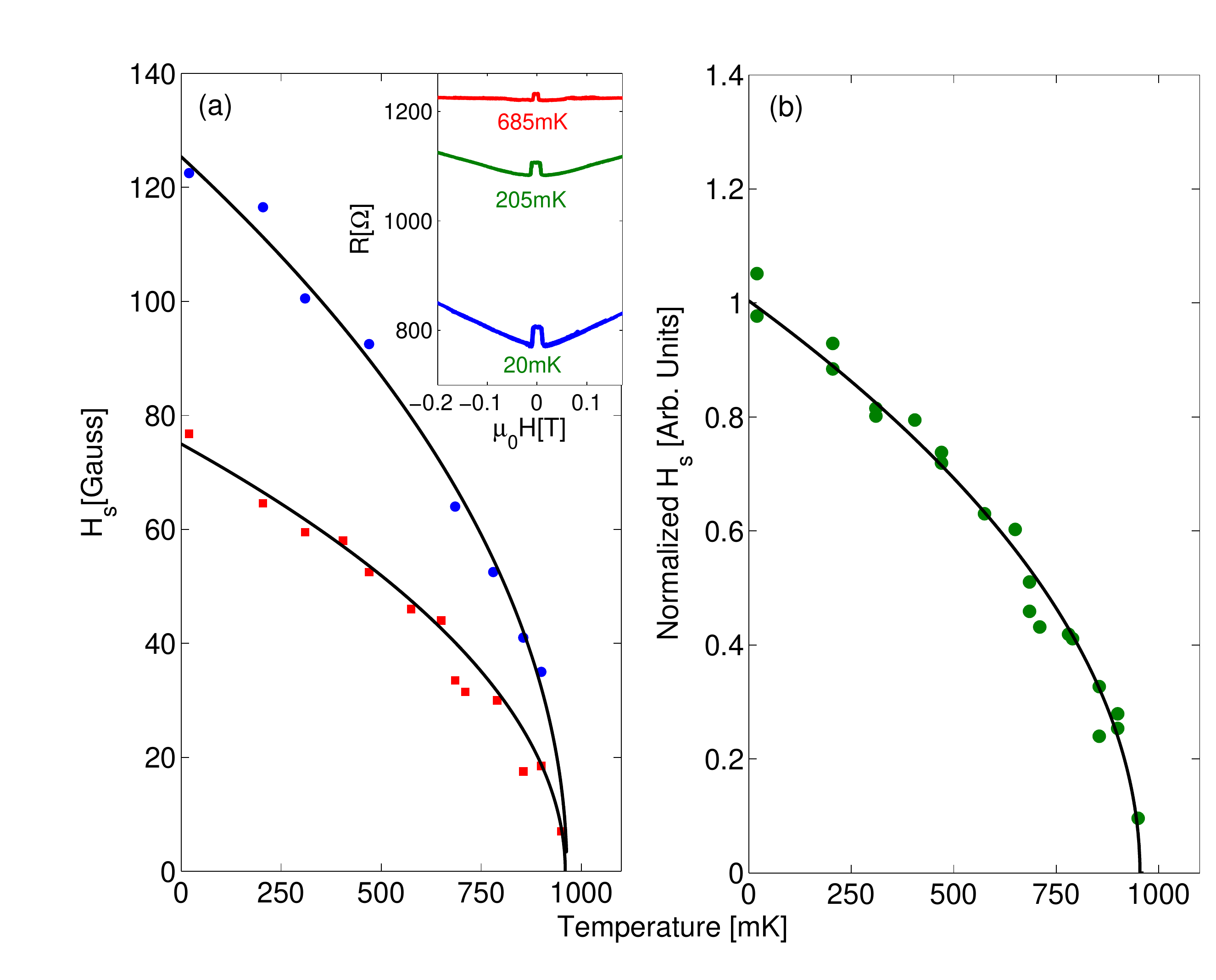}
  \caption{(color online). Inset: resistance versus magnetic field for various temperatures for parallel field orientation. (a) Saturation field $H_s$ as a function of temperature for parallel and perpendicular field orientations. The solid lines are fits to $H_s(0)(1-\frac{T}{T_{0}})^\frac{1}{2}$ (b) Normalized saturation field $\frac{H_s (T)}{H_s (0)}$ for both parallel and perpendicular orientations. ${H_s (0)}$ is determined from the fit in (a). Black solid line is a fit to $(1-\frac{T}{T_{Curie}})^\frac{1}{2}$ yielding $T_{Curie}=954 \pm 20$ mK}\label{Fig:5}
\end{figure}
\par
In the inset of Figure 5(a) we show the resistance versus magnetic field applied parallel to the nanowire. The saturation field $H_s$ is determined for each temperature. We plot the saturation field as a function of temperature for both parallel and perpendicular directions. The data fits to $H_s(0)(1-\frac{T}{T_{0}})^\frac{1}{2}$ for both orientation with $H_s(0)$ the saturation field at zero temperature. As expected $H_s(0)$ is larger for parallel field orientation comparing to the perpendicular one. The fit to the saturation field extrapolate to zero at the same temperature for both orientations within error bar.
\par
In Figure 5(b) we show the normalized saturation field $\frac{H_s (T)}{H_s (0)}$ with $H_s (0)$ being the extrapolation to zero temperature (one could use the value at 20mK with no significant difference). Both data sets collapse on the same curve. Assuming that the saturation field is proportional to the magnetization fitting the data with $(1-\frac{T}{T_{Curie}})^\frac{1}{2}$ should allow us to determine the Curie temperature to be $T_{Curie}=954 \pm 20$ mK.
\par
We note that resistance in the superconducting state is not zero (Figure 2(a)). The remnant resistance is tunable by gate voltage and is a significant fraction of the normal state value. Assuming that the sheet resistance of the nanowire for various gate voltages has similar values comparing to that obtained for larger samples it follows that the contact resistance is relatively small. Hence the residual resistance is not dominated by the contact. Finite residual resistance could arise from phase slips \cite{FINITERESISTANCENANOWIRE} due to the wire being narrower than the superconducting coherence length, $\xi\simeq100nm$ \cite{shalom2010tuning}. In order to find out the contribution of this mechanism one should study the critical exponents of the low temperature resistivity, however, in our case it is masked by the magnetic effects. Another possibility is phase separation between magnetic regions and superconducting ones.
\par
We believe that the low magnetic field effects are observable in our nano wires due to their small dimensions. This is in contrast to larger samples where the magnetic regions are shunted by low resistance, non-magnetic paths. In our nano-wires the current is forced through a series of magnetic resistors as well as through the non-magnetic (superconducting at low T) ones. Making the sample smaller than the magnetic domain allows us to set a lower limit of 50 nm (sample width) on the magnetic patch size.
\par
For field applied perpendicular to the interface (See Figure 2(e)) one expects the magnetoresistance to be smaller in the nano-wires due to the confinement as observed. In addition due to the confinement, which is not very far from the quantum limit, the phase space available for backscattering is expected to be reduced \cite{sakaki1980scattering}. We also note that the onset field for the SdH effect (See Figure 2(c)) is similar to that observed in larger samples \cite{shalom2010shubnikov}. It is possible that the SdH scattering rate (inverse Dingle time) is more sensitive to small angle scattering, which does not manifest itself in the resistivity.
\par
Various probes agree on the existence of magnetism in the \STO/\LAO~ interface but the details of the magnetic properties are still a matter of debate. Local scanning SQUID measurements, performed without the application of an external magnetic field, report the existence of randomly oriented local magnetic dipoles on the sample surface, resulting in zero or very small net magnetization over the whole sample \cite{bert2011direct}. On the other hand, torque magnetometery measurements, performed while applying an external magnetic field, revealed a non-zero sample magnetization \cite{li2011coexistence} up to room temperature. Anisotropic magnetoresistance effects were reported to persist up to a temperature of 35 K \cite{shalom2009anisotropic} \cite{flekser2012magnetotransport}. These experiments were performed at high magnetic fields, of the order of 10 Tesla, sufficient to align all spins. A similar temperature scale was also reported by the $\beta$NMR study \cite{PhysRevLett.109.257207}. It is, however, possible that while free magnetic moments do exist up to higher temperatures the magnetic order reported here forms only at lower temperatures. This view is also consistent with absence of global magnetization observed down to 1.7 K \cite{SchulerAbsenceofmagnetism}. Our experiment therefore provides the first direct indications for a non trivial magnetic ground state at the interface, which vanishes above $T_{Curie}=954 \pm 20$ mK.
\par
Finally, we would like to address the question whether superconductivity and magnetism coexist. In our case $T_{Curie}$ is of the same order as the superconducting critical temperature $ T_c$. This is a rare situation where superconductivity and magnetism can coexist. However, it is also possible that the reproducible features we observe are due to magnetic (non-superconducting) patches \cite{bert2011direct} connected in series to superconducting (non-magnetic) regions, which are spatially separated. These patches cannot be bypassed by the electric current due to the small dimensions of the bridges making magnetism visible to transport. Our experiment cannot distinguish between these two scenarios.
\par
In summary, we designed a unique method to fabricate nanowires from the \STO/\LAO~ interface. We measured transport through such 50 nm wide bridges. The Shubnikov-de Haas signal suggests that the wires have similar carrier concentration as the 2D system. For such low channel widths magnetic effects in the form of giant magnetoresistance (GMR) become visible. This GMR is found to be independent of gate voltage within the range studied. From the hysteretic behavior of resistance versus magnetic field we conclude that the moments tend to align in the plane or parallel to the crystal axis. We find that the underlaying magnetic order is not simple ferromagnetic but one with antiparallel magnetic domains. From the temperature dependence of the saturation field we find a Curie temperature of $954 \pm 20$ mK of the same order as the superconducting transition temperature.
\par
We thank A.Segal, A.Palevski, A. Gerber and N.Bachar for useful discussions. Special thanks to Glover E. Jones for help in the magnet lab. This work was supported in part by the Israeli Science Foundation under grant no.569/13 by the Ministry of Science and Technology under contract 3-8667 and by the US-Israel bi-national science foundation (BSF) under grant 2010140. A portion of this work was performed at the National High Magnetic Field Laboratory, which is supported by National Science Foundation Cooperative Agreement No. DMR-0654118, the State of Florida, and the U.S. Department of Energy. J-HP and DG acknowledge support from the Department of Energy (DOE) from grant - DOE NNSA DE-NA0001979

\bibliographystyle{apsrev}
\bibliography{House_bib}

\end{document}